\documentclass[traditabstract]{aa}  
\usepackage{graphicx}
\usepackage{txfonts}

\newcommand{\kms}[0]{\mbox{km\,s$^{-1}$}}
\newcommand{\dgr}[1]{$#1^{\circ}$}

\newcommand{\co}[0]{CoRoT}
\newcommand{\cone}[0]{\co~105895502}

\begin{document} 

   \title{Spot evolution in the eclipsing binary \cone}
   \subtitle{}
   \author{S. Czesla
          \and
          S. Terzenbach
          \and
          R. Wichmann
          \and 
          J.H.M.M. Schmitt}
   \institute{Hamburger Sternwarte, Universit\"at Hamburg, Gojenbergsweg 112, 21029 Hamburg, Germany\\
              \email{stefan.czesla@hs.uni-hamburg.de}}
   \date{Received ; accepted }

  \abstract
     {Stellar activity is ubiquitous in late-type stars. The special geometry of eclipsing binary systems
   is particularly advantageous to study the stellar surfaces and activity. 
      We present a detailed study of the 145~d \co\ light curve of the short-period ($2.17$~d) eclipsing binary \cone. 
   By means of light-curve modeling with \texttt{Nightfall}, we determine the orbital period, effective temperature,
   Roche-lobe filling factors, mass ratio, and orbital inclination of \cone\ and analyze
   the temporal behavior of starspots in the system. Our analysis shows one comparably short-lived ($\approx 40$~d) starspot, remaining
   quasi-stationary in the binary frame, and one starspot showing prograde motion at a rate of $2.3^{\circ}$~per day, whose
   lifetime exceeds the duration of the observation. In the \co\ band, starspots account for as much as $0.6$\,\% of the quadrature flux of \cone,
   however we cannot attribute the spots to
   individual binary components with certainty. Our findings can be explained by differential rotation, asynchronous stellar rotation, or
   systematic spot evolution.       
   }
  

   \maketitle

\section{Introduction}

Binary stars are numerous in the Galaxy \citep[e.g.,][]{Raghavan2010,
Yuan2015}. Among the binaries, eclipsing systems are particularly valuable
targets because they allow to study a plethora of stellar properties such as
effective temperatures, masses, radii, and limb darkening parameters 
\citep[e.g.,][]{Kjurkchieva2016}. For a number of prominent eclipsing
binaries such as AR~Lac, observational records have been accumulated reaching back more
than $100$~yr  \citep[e.g.,][]{Siviero2006}. 
Consequently, eclipsing binaries are ideal targets to put
theories of stellar structure and evolution to the test
\citep[e.g.,][]{Ribas2006, Parsons2018}.

In close binary systems, stellar rotation and orbital motion are expected to become
synchronized by tidal interaction \citep{Zahn1989}. As discussed in a review
by \citet{Mazeh2008}, observations of late-type stars in the pre-main sequence
stage by \citet{Marilli2007} and in young open clusters by 
\citet{Meibom2006} show tidal synchronization for orbital periods of less than 
10 days already at a few hundred millions years. Consequently, high
rotation rates are maintained in such systems. According to the
rotation--activity paradigm \citep[cf.,][]{Pizzolato2003}, this entails high
activity levels, of which starspots are a prominent manifestation.
Starspots can be used to study stellar rotation and its latitudinal
dependence using spectroscopy \citep[e.g.,][]{Kovari2017}, photometry
\citep[e.g.,][]{Huber2010, Nagel2016, Santos2017}, or both.

The space-based photometry provided by the CoRoT
\citep[Convection, Rotation, and planetary Transits,][]{Auvergne2009} and Kepler
satellites comprises a rich reservoir of eclipsing binary light curves
\citep[e.g.,][]{Matson2016}. In contrast to ground-based
data, neither the day--night cycle nor atmospheric turbulence interfere with
the observations of these satellites, which allows for high-quality, uninterrupted
photometric time series to be obtained, covering many months and years.

\begin{table}
\centering
\caption{Parameters of \cone}
\begin{tabular}{l l l} 
\hline\hline
Parameter & Unit & Value\\ \hline
RA (J2000)$^a$ & h:m:s & 18 43 11.074 \\
DEC (J2000)$^a$ & d:m:s & +06 11 44.779 \\
B$^b$ & mag & 13.3 \\
V$^b$ & mag & 12.6 \\
R$^b$ & mag & 12.3 \\
Distance$^a$ & pc & $680 \pm 20$ \\
Effective temperature$^a$  & K & 5440\\ \hline
\end{tabular}
\label{tab:exodat}
\tablefoot{$^a$\, Gaia DR~2 \citep{Gaia2016, Gaia2018}; $^b$\, Exo-Dat data base \citep{ExoDat2009}}
\end{table}

\section{Target, observation, and data reduction}

\cone\ is an active eclipsing binary system (2MASS J18431107+0611448, Gaia DR2 4285571561155175424).
Based on ground-based observations with the Berlin Exoplanet Search Telescope~II
(BEST II) located in Chile, \citet{Kabath2009} classified \cone\ as 
an eclipsing binary system of Algol type.
Further information on the system is given
in Table~\ref{tab:exodat}.
We here adopt the effective temperature estimate provided by Gaia. The value is probably
affected by the binary nature of the target, which is not resolved. For main sequence stars,
the quoted effective temperature corresponds to a late G spectral type with a mass of about
$0.92$~M$\odot$ \citep{Pecaut2013}. 

\begin{figure*}
\centering
\includegraphics[width=0.99\textwidth]{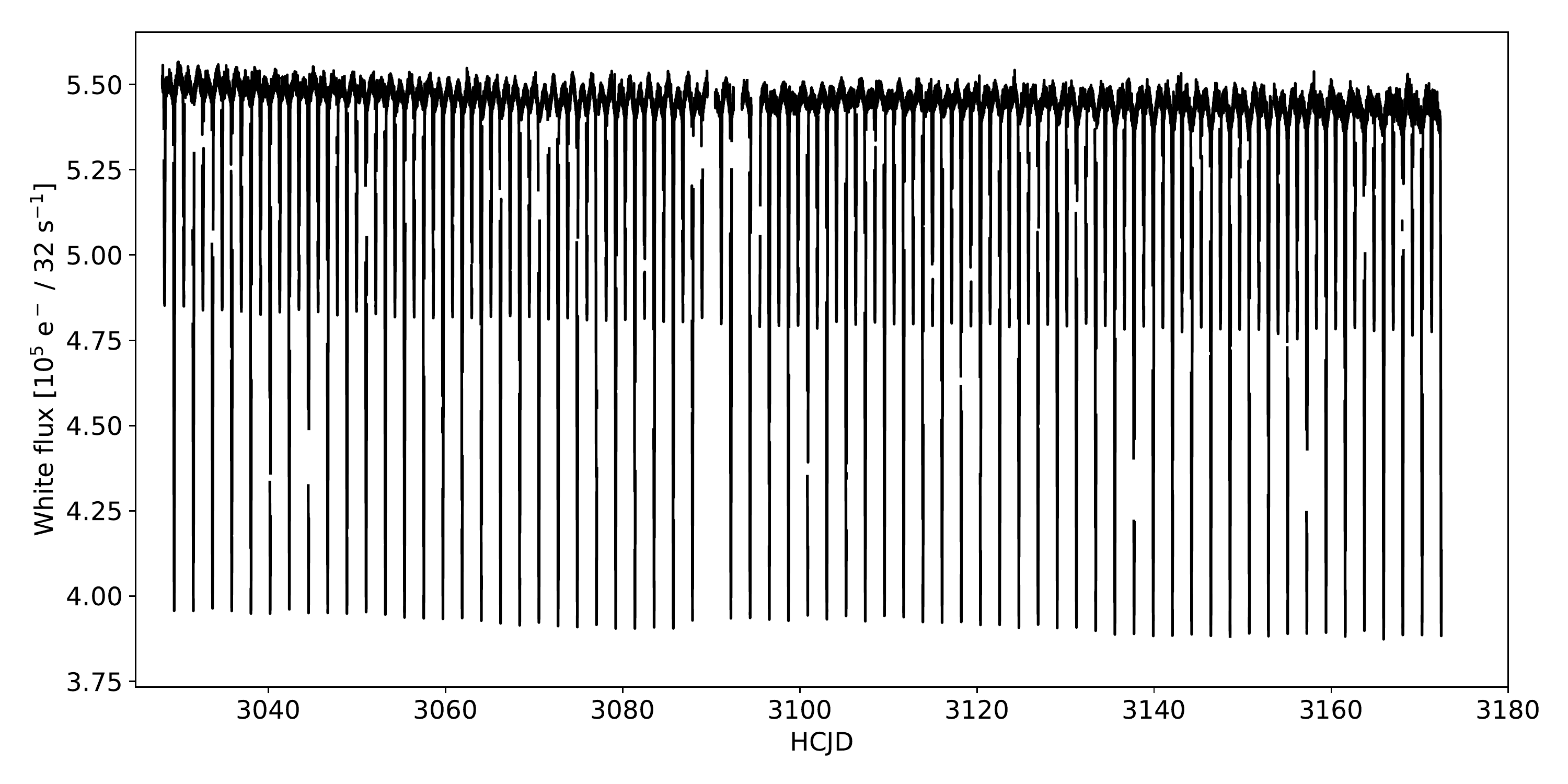}
\caption{Revised white light curve of \cone.}
\label{fig:whitetot}
\end{figure*}

The \co\ light curve of \cone\ was obtained between Mar~31 and
Sept~8, 2008, in the context of the second long run targeting the Galactic
Center (LRc02) and spans approximately 145~d.
The light curve shows strong eclipses of alternating depth with a shallower secondary eclipse
being indicative of a cooler secondary star with lower surface brightness.  
The light curve shows variable out-of-eclipse modulation attributable
to ellipsoidal variations and an evolving starspot pattern (see Fig.~\ref{fig:whitetot}).
\co\ obtained a
second light curve in July 2011, covering only 5~d.
Both light curves were observed with a temporal sampling of 32~s.
We downloaded the data from the public CoRoT~N2 data
archive. As the short 2011 light curve is only used in the
determination of the ephemeris in our analysis, we focus on the long-run light
curve in the following.

A bi-prism installed in front of the exoplanet CCD provides three
color bands dubbed red, green, and blue. However, their spectral
coverage remains uncalibrated, and therefore in this study we rely on the
so-called white flux, that is, the sum of the three color channels.
Further, we only accept  measurements marked as valid by the standard pipeline
(\texttt {STATUS} $ = 0$), which causes the conspicuous lack of
data around 3090~HCJD\footnote{Heliocentric \co\ Julian Date with origin 1 January 2000, 12:00}
(see Fig.~\ref{fig:whitetot}). Factors leading to the invalidation of flux measurements comprise the detection of
cosmic particle impacts, detector hot pixels, and crossing of the South Atlantic Anomaly.
Because the resulting light
curve shows a number of suspicious, isolated upward outliers, we
carry out an outlier rejection. In particular, we calculate the difference
between consecutive data points and disregard those showing an upward jump
exceeding 15~times the standard deviation of the noise compared to their
predecessor. With this procedure, we reject a total of 131 data points,
corresponding to only about $0.04$\,\% of the data set.
Notably, we find no
downward outliers in the light curve.

\begin{figure}
\centering
\includegraphics[width=0.49\textwidth]{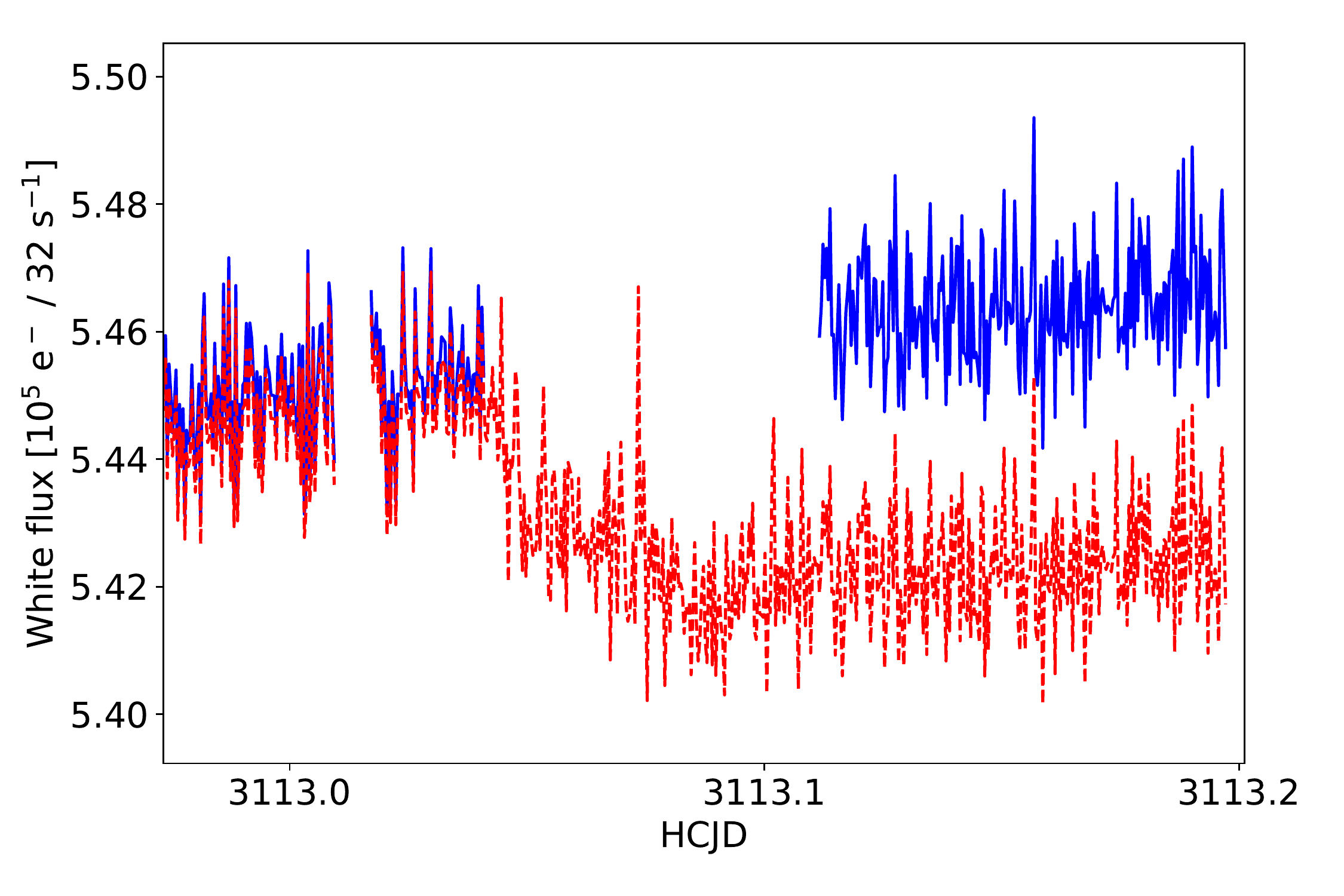}
\caption{Excerpt of the light curve around a jump before correction (red, dashed) and the
jump-corrected light curve (blue, solid) after flux level adjustment and removal of likely
affected data points.}
\label{fig:corr}
\end{figure}

After these reduction steps,
the light curve still exhibits some conspicuous, fast changes in the flux (jumps),
which we consider instrumental rather than physical in nature. By visual
inspection, we identified four such jumps, which occur
at about 3031.385, 3087.573, 3102.532, and 3113.05~HCJD. 
To correct these jumps, we take advantage of the periodic nature of the light-curve modulation, showing only relatively weak variation between
consecutive epochs (i.e., binary revolutions). Specifically, we remove the data points likely
affected by the jump process and adjust the level of the
following light-curve segment by multiplication with a factor,
determined by considering the flux ratio between the pre- and post-jump phases observed
during the preceding and following epochs.
An example of the correction is shown in Fig.~\ref{fig:corr}, and the
entire revised white-light curve is shown in Fig.~\ref{fig:whitetot}.

\section{Properties of the light curve}

The light curve displayed in Fig.~\ref{fig:whitetot} shows periodic
variability with distinct transits. With a depth of about $28$\,\%, the primary
eclipse is more than twice as deep as the secondary eclipse during which the
observed flux drops by about 11\,\%, which is indicative of a cooler secondary
component. 

\subsection{Orbital ephemeris}
\label{sec:ephemeris}

We determine the orbital period and a
reference time for the eclipses by fitting the central one-hour section of both the
primary and secondary eclipse light curves using a parabola
and estimate the instant of minimum flux based on this model.
Specifically, we use a by-eye estimate of the ephemeris as a starting solution,
determine a preliminary instant of minimum flux by a fit to the data points no
further than half an hour from the guessed position, and finally repeat the
process using the preliminary position as a starting solution.
With this procedure, we obtain the instants of minimum flux 
as a function of orbital revolution or, equivalently, epoch with respect to the
reference time.

\begin{figure}[h]
    \includegraphics[width=0.49\textwidth]{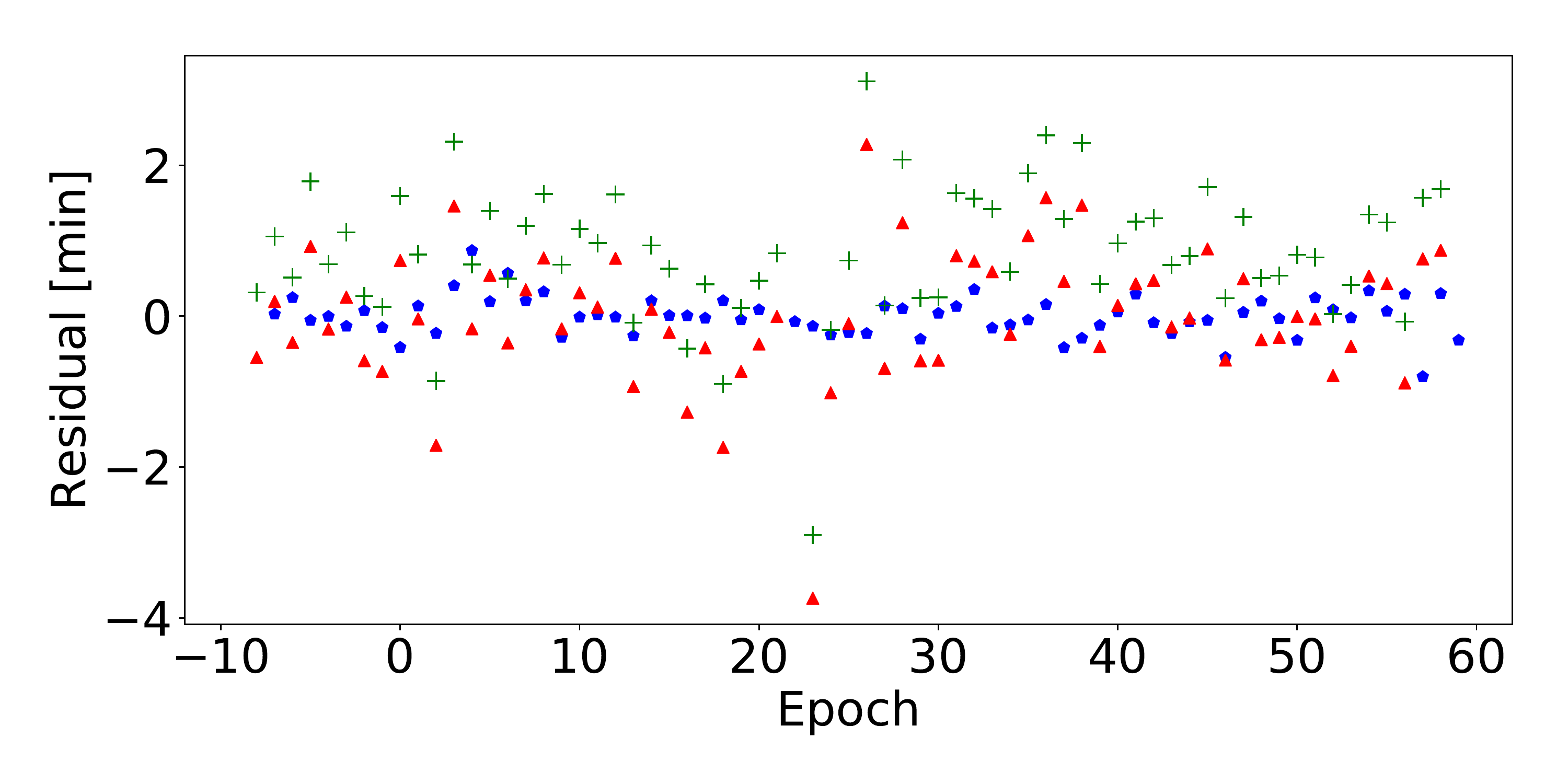}
    \caption{Residuals of estimated eclipse times with respect to linear
    ephemeris for primary (blue dots) and secondary (red triangles) eclipses.
    Green crosses indicate the residuals of the secondary eclipse times with
    respect to the ephemeris derived from the primary eclipse timing.
    \label{fig:ephRes}}
\end{figure}

We next determine the orbital period using a linear fit to the instants of
primary eclipse. The resulting period and reference time
are given in Table~\ref{tab:ephemeris} and the residuals 
are shown in Fig.~\ref{fig:ephRes}; note that two additional primary
eclipses as well as one secondary eclipse at epochs 533 and 534 constrain
the ephemeris but are not shown here. The error was estimated using the
jackknife procedure, which is a resampling technique \citep{Efron1981}.
To estimate jackknife errors, we repeat the analysis for all subsamples of
data, which can be obtained by removing a single data point from the original set.
The uncertainty of the best-fit value can then be estimated
from the width of the thus-obtained distribution of parameter estimates.

While the secondary and primary eclipses are expected to show identical
periods, the phase offset is not necessarily $0.5$ if the binary orbit is not
circular. In Fig.~\ref{fig:ephRes}, we also show the residuals of the
secondary eclipse times with respect to the primary ephemeris, which show a
systematic offset compared to the nominal
value of 0.5 valid for a circular orbit.
Therefore, we set up a model with three free parameters, namely the reference
time, the orbital period, and the phase offset of the secondary eclipse, and fit
the parameters to the primary and secondary eclipse times. The resulting values are
given in Table~\ref{tab:ephemeris}.
While reference time and orbital period are consistent with those determined
from the primary eclipse timing to within the uncertainty, a significant phase
offset is found, corresponding to an average lag of $49\pm 4$~s in the secondary
eclipse time with respect to the prediction of a circular orbit; uncertainties
are again derived using the jackknife.

The scatter in the residuals corresponding to the secondary eclipse times is
larger and also appears to show systematic variation
around epoch~20. While the larger scatter may be related to the smaller
curvature of the secondary eclipse light curve compared to the center of the
primary eclipse, the systematic variation is probably attributable to
occulted starspots on the secondary component, making the timing of the primary
eclipses more reliable.
In the following analysis, we rely on the ephemeris derived from the combined
model with a phase offset.

Based on the delay of the secondary eclipse, a value of $4\times
10^{-4}$ is obtained for the combination of eccentricity and argument of
periastron, $e\cos(\omega)$, which is also a lower limit for the eccentricity
\citep[e.g.,][]{Matson2016}. Owing to the ellipsoidal variation and the effect of starspots,
a difference in the duration of the eclipses is hard to determine accurately.
Assuming a conservative by-eye upper limit of $600$~s, an eccentricity in excess
of $1.8$\,\% can be ruled out however. We note that a fraction of the delay
will be contributed by the light travel time effect \citep[e.g.,][]{Kaplan2010},
causing a delay, $\Delta t_{\rm LT}$, of
\begin{equation}
  \Delta t_{\rm LT} = \frac{P_{\rm orb} K_2}{\pi c} (1-q) \approx
  0.2\,\mbox{s$^2$\,km$^{-1}$} \times K_2\, (1-q) \; .
\end{equation}
Here, $q$ is the binary mass ratio and $K_2$ (in \kms) is the radial-velocity
semi-amplitude of the secondary component, both of which are unknown.
Conservatively assuming $100$~\kms\ for $K_2$ and a mass ratio of $0.5$, we
estimate $\Delta t_{\rm LT} \lessapprox 10$~s in our case.

\begin{table}
\centering
\caption{Fit results for the ephemeris
\label{tab:ephemeris}}
\begin{tabular}[h]{l l} \hline \hline
  Parameter & Value \\ \hline
  \multicolumn{2}{c}{Primary eclipse only} \\
  $T_{0, \rm pe}$ & $3044.525388\pm 0.000025$~HCJD \\   $P_{\rm pe}$ & $2.1678211 \pm 0.0000003$~d \\ \hline
  \multicolumn{2}{c}{Both eclipse with phase offset} \\
  $T_0$    &   $3044.525402\pm 0.000029$~HCJD \\   $P$      &  $2.1678209 \pm 0.0000004$~d \\
  $\Delta \phi - 0.5$ &  $0.00026 \pm 0.00004$ \\ \hline
\end{tabular}
\end{table}

\subsection{Evolution of the noise level}
To prepare the modeling, we study the noise behavior of the light curve. In particular,
we employ the $\beta\sigma$ procedure by \citet{Czesla2018}. If the signal were constant,
an estimate of the sample standard deviation of the data would immediately yield the noise level.
This is no longer true when there is considerable intrinsic variation in the signal, as in this case. 
The idea behind the $\beta\sigma$ technique
is to estimate the standard deviation of the noise, $\sigma$, by studying the distribution of numerical
derivatives of the data (the $\beta$ sample). In this way, the impact of intrinsic variation on the
noise-level estimation can be minimized.  
For example, the difference between any pair of consecutive data points can be used to construct a sample
of first-order derivatives, which will not be affected by a signal, varying only weakly between consecutive
data points.
In this case, we opted for second-order numerical derivatives and skipped every second data point in the
calculation of the derivatives to avoid potential correlation.   
Thus, we obtained the
standard deviation of the noise for all epochs comprising more than 100 data points.
Finally, we calculate the signal-to-noise ratio (S/N) by dividing the mean level of the light curve
in each epoch by the noise level.

\begin{figure}
        \includegraphics[width=0.49\textwidth]{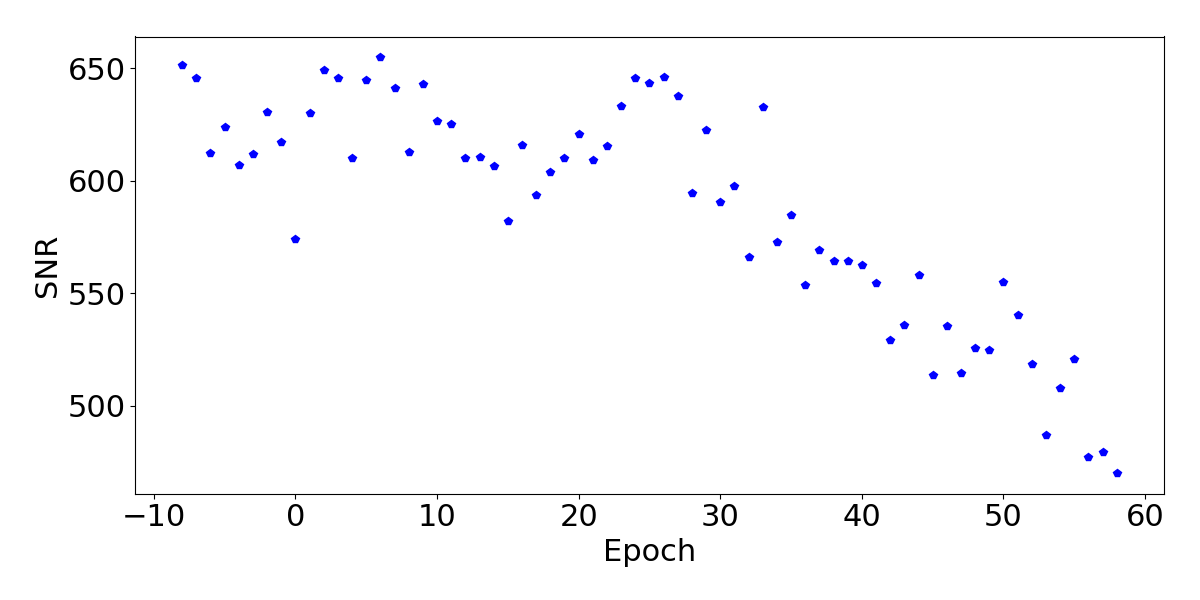}
    \caption{The S/N as a function of epoch.
    \label{fig:SNR}}
\end{figure}

The evolution of the S/N is shown in Fig.~\ref{fig:SNR}. The S/N remains relatively constant
between about $600$ and $650$
until around epoch~30. It then continuously decreases, indicating a loss of quality in the observations.

\subsection{Light curve asymmetry}
\label{sec:symmetry}

The light curve shows distinct time-variable modulation between the transits, which is
characteristic of ellipsoidal variations and an evolving starspot configuration \citep[e.g.,][]{Strassmeier2009}. 
While it seems improbable that there is any epoch without a contribution of starspots to
the light curve, their impact differs among the various observed epochs.
In view of the light-curve-modeling process, it is helpful to identify epochs for which some
information on the instantaneous starspot configuration can be derived from more fundamental
considerations. Here, we attempt to employ symmetry to identify such epochs.

We refer to two viewing geometries as \emph{equivalent} if they can be transformed into
each other by reflections or rotations in the plane of the sky. Such configurations therefore
yield the same observed flux if Doppler boosting is neglected.
Assuming a circular orbit, an aligned system, and the absence of starspots, the viewing geometry is equivalent for
all instances $T_t + \tau$ and $T_t - \tau$, where $T_t$ is the center of any primary or secondary
eclipse and $\tau$ is some time offset. It therefore follows for the observed flux, $f(t)$, that
\begin{equation}
    f(T_t + \tau) = f(T_t - \tau) \; .
\end{equation}
The above relation continues to hold when starspots are added, if these are symmetric with respect to
the plane generated by the vector connecting the stars and the orbit normal. By quantifying 
the accuracy of the relation, we can therefore identify epochs with a bona-fide symmetric starspot configuration. 

\begin{figure}
\centering
\includegraphics[width=0.49\textwidth]{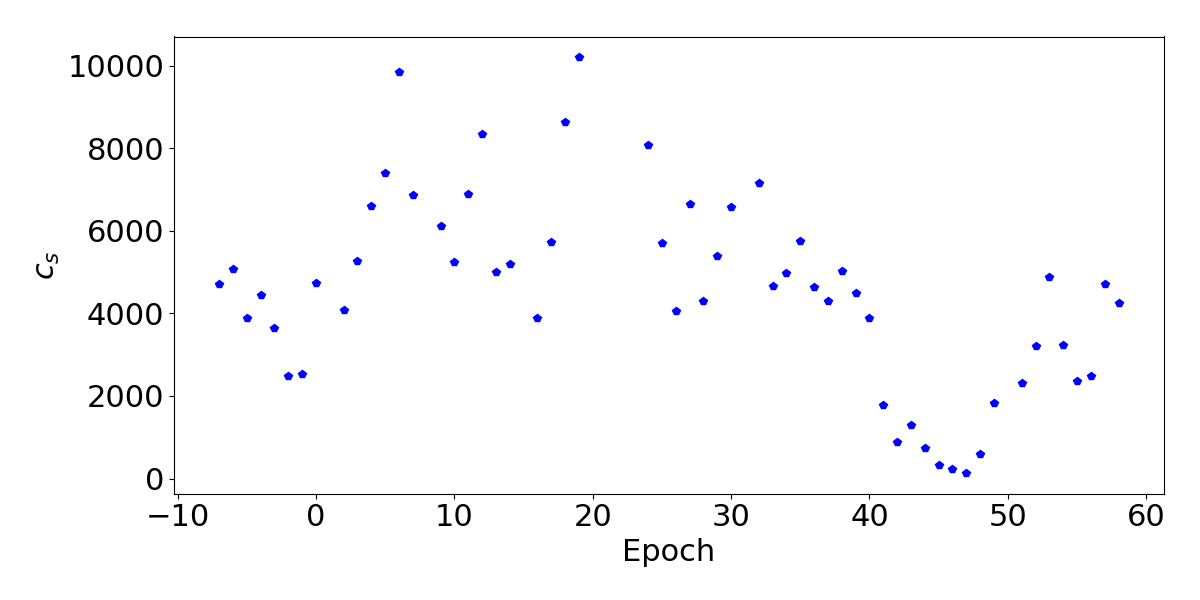}
\caption{The asymmetry coefficient as a function of epoch.}
\label{fig:Course}
\end{figure}

In practice, we went through all observed epochs and divided the orbital phase intervals $\phi_a = 0.05-0.45$
and $\phi_b = 0.55-0.95$ into 20 equidistant bins each. The phase intervals are chosen such that the eclipses are
excluded. We then defined the symmetry coefficient, $c_s$, as
\begin{equation}
    c_s = \sum_{i=1}^{20} \frac{(a_i - b_i)^2}{\sigma_{a,i}^2 + \sigma_{b,i}^2} \; ,
\end{equation}
where $a_i$ and $b_i$ denote the mean value of the data points in the i-th bin in the phase intervals $\phi_a$ and $\phi_b$
and $\sigma_i$ denotes the standard deviation of the sample mean of the data points in the respective bin.
The thus-defined symmetry coefficient
is shown as a function of epoch in Fig.~\ref{fig:Course}. 

The symmetry coefficient assumes a minimum for a symmetric light curve. According to
Fig.~\ref{fig:Course} a particularly symmetric period was observed between epochs~40 and 50, which we attribute to
a special starspot configuration.
The lowest value of the symmetry coefficient is obtained for epoch ~47 for which we find a numerical value of $135$. 
Under the null hypothesis of identical expectations in all bins $a_i$ and $b_i$ (and normal errors), $c_s$ follows
a $\chi^2$-distribution with $20$ degrees of freedom. The null can therefore be formally rejected at high significance,
even for the most symmetric configuration observed according to the above condition.

\section{Light-curve modeling}

To model the light curve, we use the open source \texttt{Nightfall} code
developed by one of the authors (R.W.). Here,
we use version 1.88 (released Nov 2015). \texttt{Nightfall} models stars
as equipotential surfaces of the Roche potential, using the geometric setup
described by \citet{Djurasevic1992}.
The bolometric correction
for mutual reflection follows the prescription by \citet{Hendry1992}.
The correction is calculated by an iterative procedure, and we use two iterations here after verifying
that this suffices for the required accuracy.
For the gravity-brightening exponent of convective stars, \texttt{Nightfall} uses 
the results from \citet{Claret2000a}, which provide a smooth transition to the
\citet{vonZeipel1924} exponent for radiative stars.
For temperatures below 9800\,K (as is the case here), model fluxes are based 
on PHOENIX models for solar abundances \citep{Hauschildt1999,Hauschildt2003}. 
Limb darkening coefficients are taken from
\citet{Claret2000b}. \texttt{Nightfall} provides tables of
pre-computed band fluxes integrated over the supported filter bandpasses and
interpolates in effective temperature $T_{\rm eff}$, but not in
surface gravity $\log{g}$, hence, we use a value of $\log{g} = 4.0$ for both
stars. 
To account for Doppler boosting, {\tt Nightfall} applies a correction
according to Eq.~3 in \citet{Kerkwijk2010}.
The actual CoRoT passband is relatively broad \citep{Auvergne2009}.
As CoRoT registers about $78$\,\% of the signal of \cone\ in its so-called red color band (which remains
uncalibrated, however), we use the R-band model fluxes to represent the light curve
in our modeling with \texttt{Nightfall}.

Nightfall can take into account circular spots on both of the binary components. The
spots are described by their stellar longitude and latitude, their radius, and a
dimming factor, which specifies the ratio between starspot and photospheric
effective temperature. To alleviate the effect of the notorious degeneracy between starspot
radius and dimming factor in the modeling, we here consider only spots with a fixed radius of
$20^{\circ}$.
By opting for a large starspot radius, we simultaneously
minimize effects related to the surface discretization on the objective function (i.e.,
$\chi^2$) in the minimization process.

\begin{figure}
    \centering
    \includegraphics[width=0.39\textwidth]{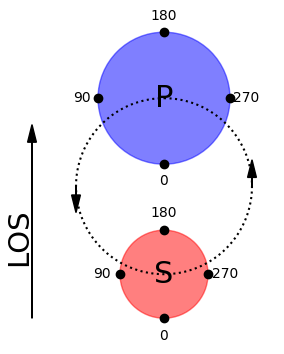}
    \caption{Convention of stellar longitude used in Nightfall.
    \label{fig:nfgeometry}}
\end{figure}

The convention of stellar
longitude adopted by Nightfall is shown in Fig.~\ref{fig:nfgeometry}.
The starspot latitude is often also rather ill-constrained in light-curve inversion
problems because it is strongly degenerated with other properties, specifically the dimming factor
and radius in our case. We therefore also fix the starspot latitude such that starspots are centered on
the stellar equator. We note that the eclipses are particularly informative phases for breaking this
degeneracy, however our modeling is focused on the inter-eclipse phases. 

The assumption of large, circular spots is a simplification adopted only for
modeling purposes. On the Sun, individual spots are typically much smaller, although they frequently appear
in groups \citep[e.g.,][]{Bogdan1988}. Although the stars studied here are considerably more active than the Sun, their surfaces are
also most likely covered by an ensemble of spots following some size distribution rather than
giant monolithic spots. However, the small-scale structure remains unresolved in our
light-curve modeling. As shown by \citet{Jeffers2005} and \citet{Ozavci2018} the model spots absorb the combined
effect of the longitudinally asymmetric part of the starspot configuration. \citet{Ozavci2018} specifically use
the term ``spot cluster'' to underline this fact. While we stay with the spot nomenclature in the following, the
approximative nature of the model should be kept in mind.

\subsection{Spot quadrature depression}
The impact of a starspot at a given position on the model light curve is determined by the dimming factor and the
spot radius. Because of the strong degeneracy, usually only some combination of these quantities is constrained in
the modeling. Even after we break this degeneracy (not resolve it of course) by fixing the radius, the interpretation
of the dimming factor in terms of its impact on the light-curve model remains somewhat intricate.
Therefore, we here define a quantity, which can serve as a more direct measure of the spot impact on
the model light curve.

When a single spot is put at a stellar longitude of \dgr{90} (and zero latitude), it faces the observer
at quadrature between secondary and primary eclipse. Here, we define
the {spot quadrature depression} (SQD) as the relative loss of flux caused by a spot 
of a given dimming factor and radius, located at a stellar longitude of \dgr{90}, during quadrature phase compared to
an unspotted model. The SQD is a measure of the spot impact on the model, combining the effects of
the dimming factor and the radius. It is also comparable for
spots on the primary and secondary component, whose effective temperatures differ, further complicating the
interpretation of the dimming factors.

\subsection{The fundamental binary parameters}
\label{sec:fundamental}

Our goal here is to determine
the fundamental system parameters, namely the orbit inclination ($i_{\rm orb}$), the mass ratio ($q$), the
Roche lobe filling factors for primary and secondary ($f_{\rm P}$ and $f_{\rm S}$), and the effective temperatures
($T_{\rm eff,P}$ and $T_{\rm eff,S}$), which we consider constant throughout
the observed light curve. In addition, we allow for a free normalization factor in the model to represent the unknown absolute flux level,
which we consider a nuisance parameter in the modeling. 
As the light-curve modeling is mainly sensitive to the ratio of surface brightnesses in the observed band, we
fix the effective temperature, $T_{\rm eff,P}$, of the primary star to $5500$~K in accordance with
the Gaia results (Table~\ref{tab:exodat}) and further assume a value of four for $\log(g)$ for both components
of the system.

Time-dependent starspot patterns represent a nuisance in the context of determining the binary parameters,
because they add more free parameters to the problem.
In Sect.~\ref{sec:symmetry}, we concluded that it is unlikely that the light curve is free
from starspot contributions during any observed epoch but found that epoch~47 displays the most symmetric
light curve. While the true number of starspots remains unknown, we are confident that
they were located close to the central plane during this epoch and that their combined effect can therefore
be absorbed by a single circular spot in the modeling.   

\begin{figure}
\centering
\includegraphics[width=0.49\textwidth]{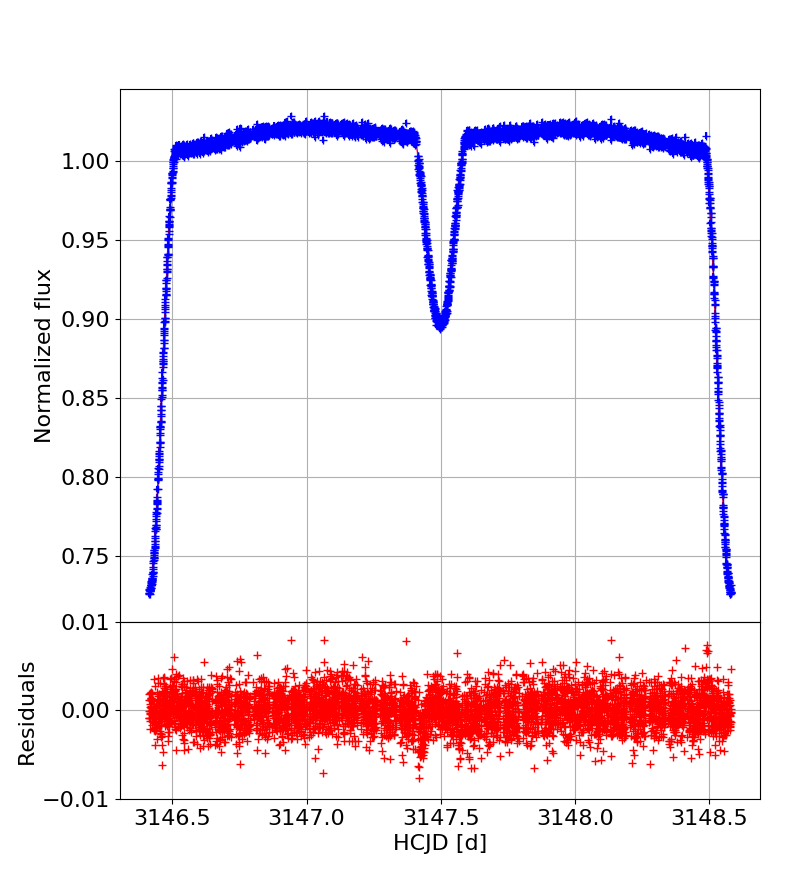}
\caption{Light curve (top) along with the residuals with respect to the best-fit model for epoch 47 (bottom).}
\label{fig:fit1}
\end{figure}

In Fig.~\ref{fig:fit1} we show the observed CoRoT light curve during epoch~47 along with our best-fit model and
the residuals. The best-fit parameters are listed in Table~\ref{tab:fit1}. Formally, we obtain a $\chi^2$ value of
$5075.8$ with 5158~degrees of freedom and, consequently, a reduced $\chi^2$ value of $0.98$. Nonetheless, some systematic
residuals can be identified, which are associated with the beginning of ingress and the end of egress. This may be attributable
to the adopted description of limb-darkening or unaccounted-for contributions of potentially occulted starspots.
To test for
the randomness of the residuals, we carried out
a Wald-Wolfowitz runs test on their signs \citep[][]{Wald1940, Bradley1968}. In particular, we consider here any
consecutive sequence of two or more residuals with the same sign a ``run''. We  
find a total of 1283 such runs in the residuals of the fit. Under the null hypothesis of
independent residuals with equal expectation of being positive or negative, we find an expected number of $1291 \pm 18$ runs
by means of a simulation, which provides no evidence against the null. Therefore, we find
both the $\chi^2$ and the runs test to be consistent with an acceptable fit result.

\begin{figure}
\centering
\includegraphics[width=0.49\textwidth]{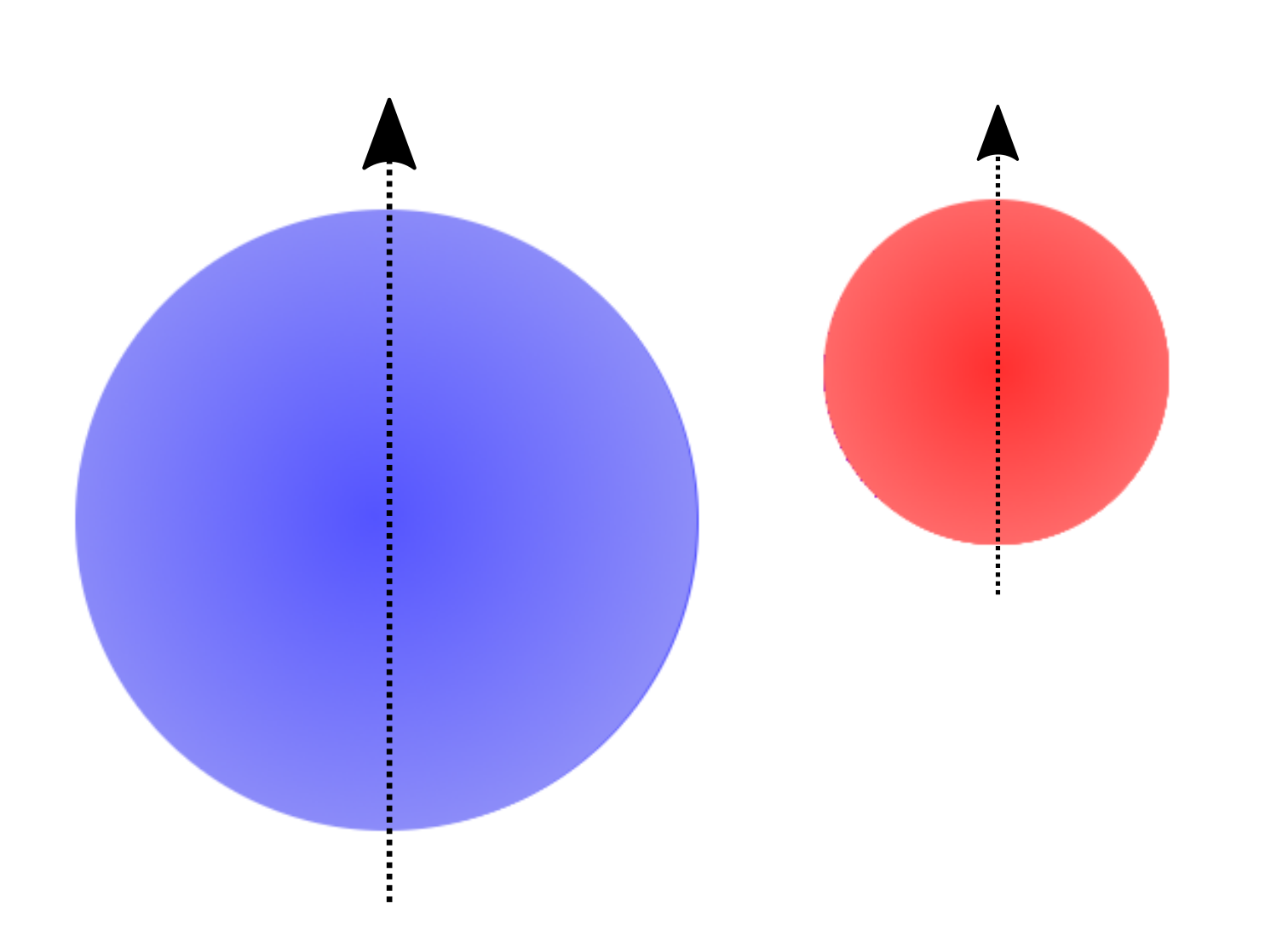}
\caption{System geometry with primary (blue, left) and secondary (red, right) at orbital phase 0.44, i.e., shortly before
secondary eclipse. Dotted arrows mark the model stellar rotation axes, which are assumed to be aligned with the orbit normal.}
\label{fig:system}
\end{figure}

To verify that a model without a starspot does not yield an equally acceptable result,
we repeat the minimization without a starspot, which increases the number of degrees of freedom by two.
With this approach, however, we end up with a best-fit reduced $\chi^2$ value of $2.1$. By means of an F-test, we
conclude that the one-spot model provides a superior fit at a significance level of $8.3$\,$\sigma$, leaving little room for controversy
regarding this issue. In Fig.~\ref{fig:system} we show the system geometry as seen from Earth,
resulting from our modeling. The geometry is such
that the secondary eclipse is almost total.  

In Table~\ref{tab:fit1}, we also provide a statistical error estimate, derived from a
Markov-Chain Monte-Carlo (MCMC) analysis of the posterior probability distribution. The uncertainties are approximated
by the standard deviation of the respective marginal distribution. We caution, however, that the posterior and therefore these uncertainties
are all conditional on the assumptions, such as that of large, circular spots on the equator.
This is most obvious in an uncertainty of $0.1$~K for $T_{\rm eff, S}$,
which obviously hinges on the fixing of $T_{\rm eff,P}$. To study the dependence of the system parameters on the assumed
primary effective temperature, we repeat the modeling with values between $5000$~K and $7000$~K. The only parameters
strongly affected are the effective temperature of the secondary and the spot-dimming factor. In the considered range,
the change in these
parameters is well reproduced by linear relations for which we find slopes of
\begin{equation}
    \frac{\partial \mbox{$T_{\rm eff,S}$}}{\partial \mbox{$T_{\rm eff,P}$}} = 0.64 \;\;\; \mbox{and} \;\;\;
    \frac{\partial \mbox{$d_{\rm Spot}$} }{\partial \mbox{$T_{\rm eff,P}$}} = -3.55 \times 10^{-5}~\mbox{K$^{-1}$} \; ,
\end{equation}
where $d_{\rm Spot}$ is the spot dimming factor. The inherent uncertainty in $T_{\rm eff,P}$  therefore does not severely impede our
ability to determine the system geometry. Nonetheless, systematic uncertainties likely also affect other parameters such as the mass ratio, $q$,
which is notoriously difficult to determine via light-curve modeling in detached binary stars \citep[e.g.,][]{Wilson1994}.
In the following, we use the here-obtained best-fit model to represent the binary
light curve without starspot contributions in the following modeling.

\begin{table}
\centering
\caption{Best-fit (minimum $\chi^2$) parameters for CoRoT~105895502 derived from the light curve analysis of epoch 47.
\label{tab:fit1}}
\begin{tabular}{lll} 
\hline\hline
Parameter & Value & Statistical  \\ 
& & uncertainty  \\ \hline
q & 0.426 & 0.004 \\
$i_{\rm orb}$ [$^{\circ}$]  & 84.15 & 0.02 \\
$T_{\rm eff,P}$ [K] & 5500  & fixed \\
$T_{\rm eff,S}$ [K] & 4639 & 0.1 \\
$f_{\rm P}$ & 0.4454 &  0.0008  \\
$f_{\rm S}$ & 0.3700 & 0.001 \\
 & & \\
lon$_{\rm spot}$ [$^\circ$] & 363.7 & 0.4  \\
dimming & 0.912 & 0.002 \\
\hline
\end{tabular}
\end{table}

\subsection{One- and two-spot models for all epochs}

A model with a single starspot as introduced in Sect.~\ref{sec:fundamental} yields an
acceptable fit for the particularly symmetric light curve of epoch~47. We now
use a similar approach to model the light curve of all available epochs. Here we focus,
however, on the starspot evolution and assume the light curve of the unspotted binary to be known.
Specifically, we
fix all parameters unrelated to the spot properties
to the best-fit values derived in Sect.~\ref{sec:fundamental}.
In the following modeling, we consider the fits to all individual epochs independent.   

\subsubsection{One-spot modeling}

In a first attempt, we fit the light curves of all individual epochs assuming a single starspot
with a radius of $20^{\circ}$ located on the equator of the secondary component. In the fit, the spot
longitude and dimming factor as well as the model normalization are considered free parameters.  

\begin{figure}
    \includegraphics[width=0.49\textwidth]{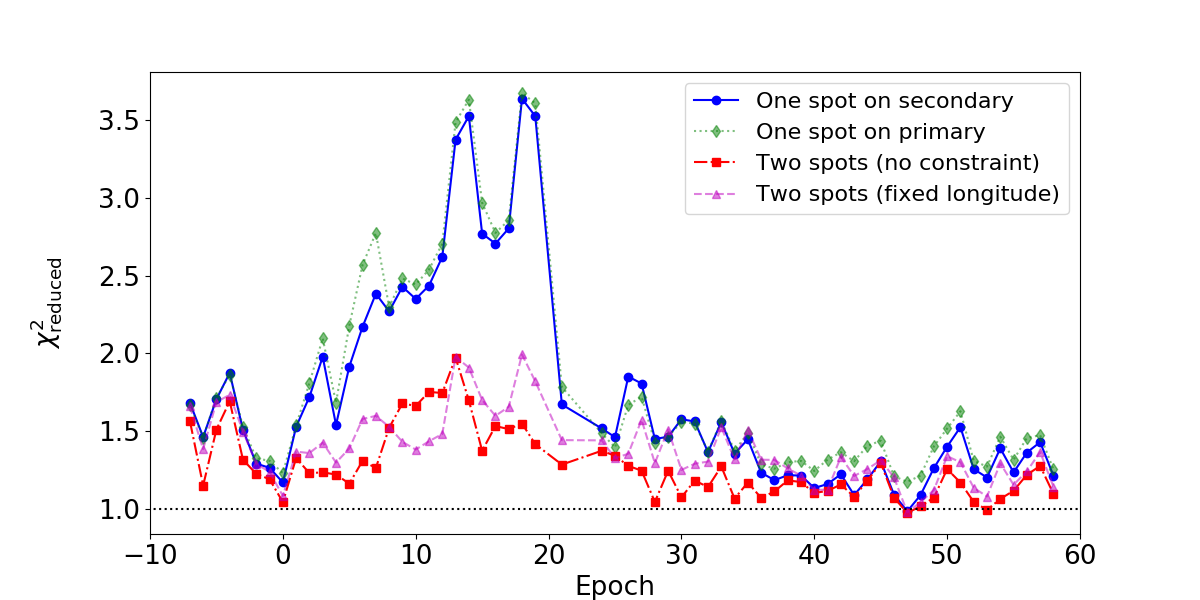}
    \caption{Evolution of the reduced $\chi^2$ value as a function of epoch for the one-spot (blue circles) and
    two-spot models (red stars). 
    \label{fig:chievo}}
\end{figure}

In Fig.~\ref{fig:chievo}, we show the temporal evolution of the resulting reduced $\chi^2$ value. Around epoch~47
the fit quality is optimal, which we attribute to the fact that the binary parameters were derived based on this epoch. 
Between about epoch~0 and 20, the fit quality of the one-spot model is worse than during the rest of the light curve;
we postpone a more detailed discussion of this fact. We note that the fit quality is essentially identical if the spot is
located on the primary instead of the secondary component (see Fig.~\ref{fig:chievo}).

\begin{figure}
    \includegraphics[width=0.49\textwidth]{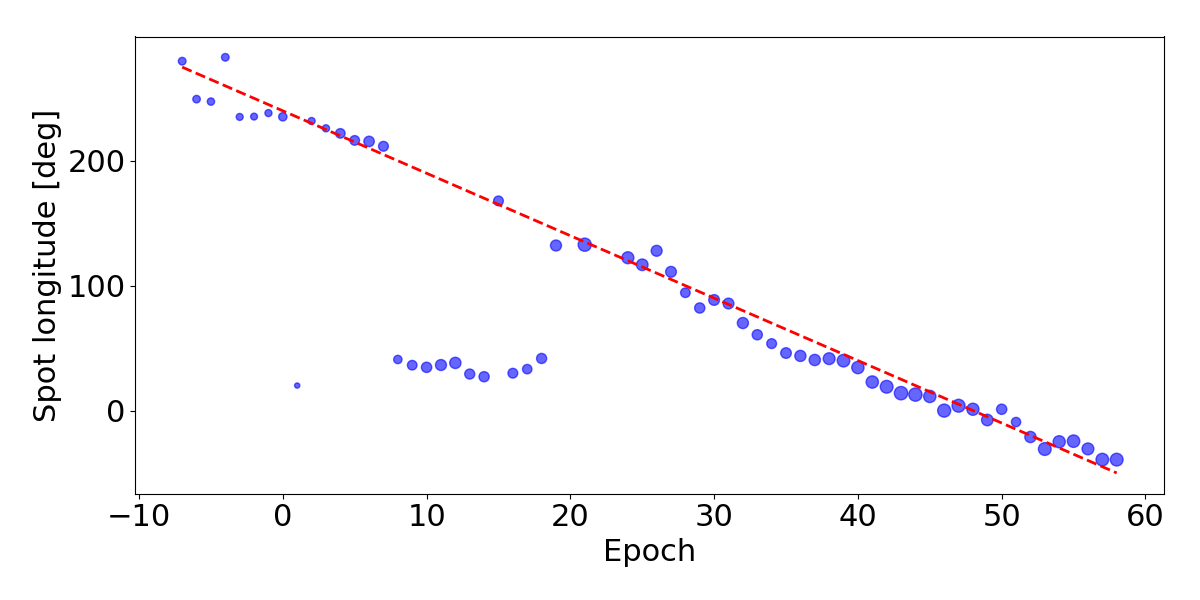}\\
    \includegraphics[width=0.49\textwidth]{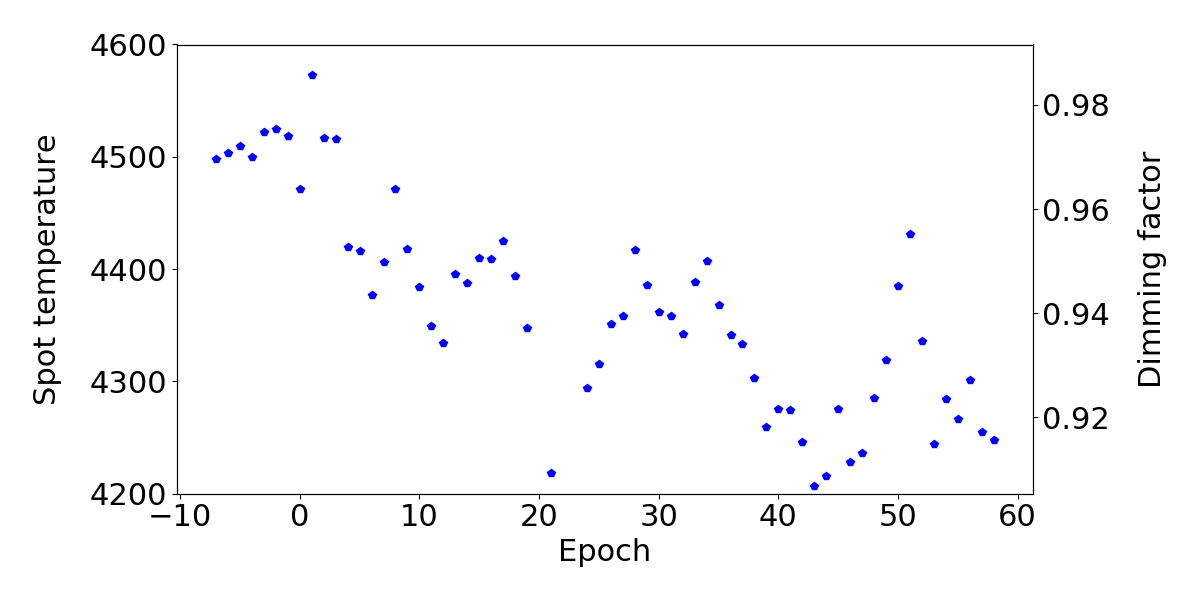}
    \caption{Top: Time evolution of spot longitude based on one-spot model along with an approximation to the observed trend (dashed red line).
    Symbol size is proportional to spot temperature contrast. Bottom: Time evolution of spot temperature in model.    
    \label{fig:onespot}}
\end{figure}

In the top panel of Fig.~\ref{fig:onespot}, we show the evolution of spot longitude as a function of epoch. With the exception
of the range between epochs 0 and 20, the time evolution of the spot longitude, $l_{\rm Spot}$, is well described by a linear model of the form
\begin{equation}
    l_{\rm Spot} = 240^{\circ} - 5^{\circ} \times E \; ,
    \label{eq:lon}
\end{equation}
where $E$ is the epoch. According to our model, the spot shifts by about $5^{\circ}$ per epoch or about $2.3^{\circ}$ per day.
With respect to the binary orbit, the spot advances in the direction of rotation, that is, its motion is prograde. Based on
this relation, we estimate an uncertainty of about $4^{\circ}$ for {individual} spot longitudes. The scatter is
larger at the beginning until epoch~0, where the spot temperature contrast is also lower.
During the observed time span,
the model spot almost completes a revolution. Meanwhile, also its dimming evolves. The spot dimming increases most strongly
until about epoch~20 after which further dimming
appears to slow or eventually stop; we recall here, however, the worsened fit quality between epochs 0 and 20.
Extrapolating the spot evolution into the domain prior to the start of the observation, we speculate that the spot emerged
about ten epochs before the observation commenced. At any case, the spot lifetime is longer than the observational domain of
$145$~d.

\subsubsection{Two-spot modeling}

The fit quality of the one-spot model as shown in Fig.~\ref{fig:chievo} is worst for the light curves of epochs~0 to 20,
indicating that the model is insufficient to describe the
data there. Therefore, we introduce a second spot in our modeling. As for the first spot, we fix the
spot latitude and radius and only allow the longitude and dimming factor to vary. In order to avoid ``spot collisions''
in the modeling, we place this second spot on the primary component of the binary, which is not necessarily its physical
location however. In the following, we refer to the spots as the primary and secondary spot depending on the component
on which they reside.

In a first attempt at the two-spot model, we vary the longitudes and dimming factors of both spots along with the normalizing
constant. The resulting fit quality is shown in Fig.~\ref{fig:chievo} (red squares). Clearly, the fit quality for the light curves
between epochs~0 and 20 improves substantially owing to the introduction of the primary spot, and
expectedly the fit is not worsened otherwise.
This justifies the introduction of
the primary spot at least during the said epochs, for which a substantial improvement in the fit quality is obtained.
Still, a small
decrease in fit quality persists between epochs~0 and 20, which may be attributable to the presence of further
spots or the limitations of our spot representation in the modeling.

\begin{figure}
    \includegraphics[width=0.49\textwidth]{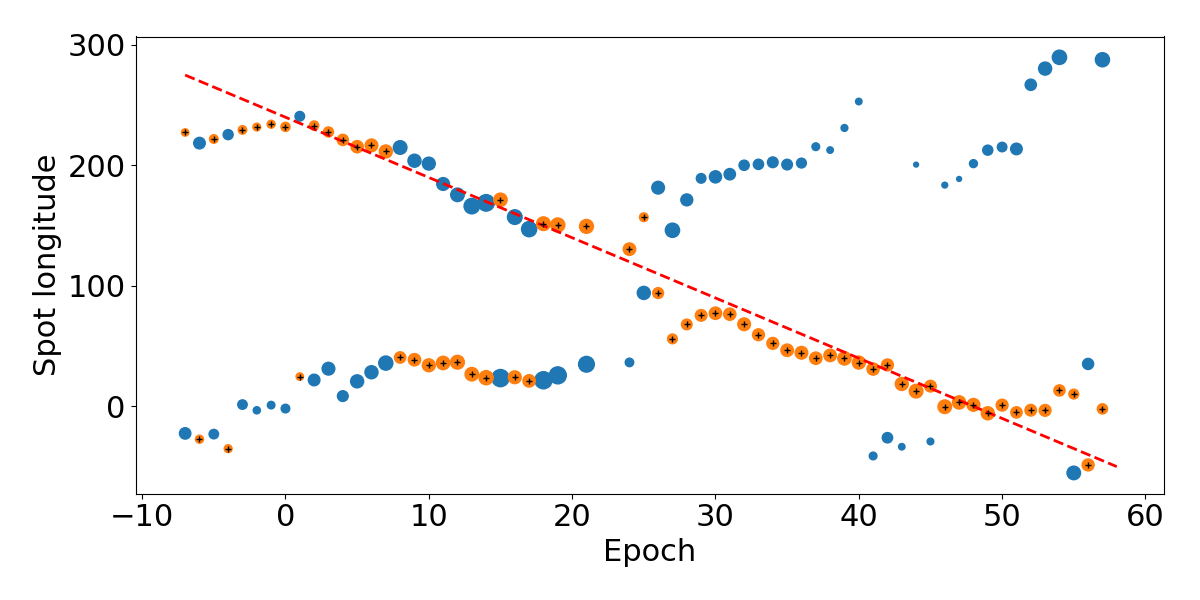}\\
    \includegraphics[width=0.49\textwidth]{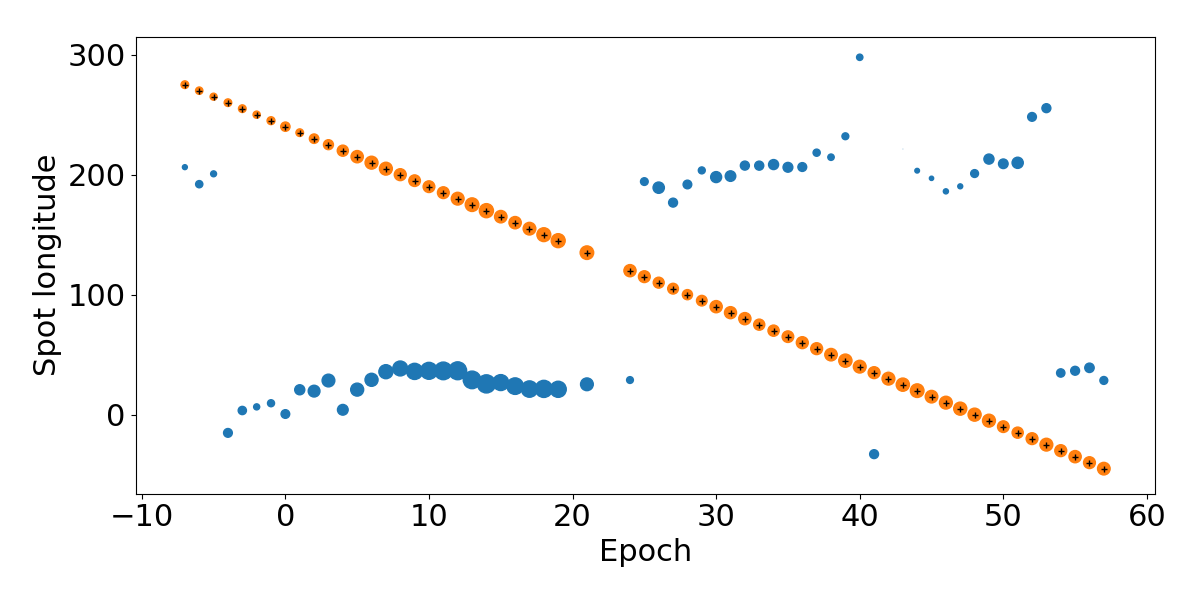}
    \caption{Time evolution of spot longitudes in the two-spot model. The primary and secondary spots are indicated by 
    orange circles and blue circles, respectively; an additional cross marks the secondary spot.
    Upper panel: Model without longitudinal constraints. Evolution
    from one-spot model (Eq.~\ref{eq:lon}) as dashed, red line. Bottom panel: Two-spot model with secondary spot longitude
    fixed according to Eq.~\ref{eq:lon}. 
    \label{fig:twospotevo}}
\end{figure}

The top panel of Fig.~\ref{fig:twospotevo} shows the time evolution of the longitudes
of both spots, again encoding the dimming factor by symbol size and superimposing the linear trend derived from our previous one-spot modeling.
Clearly, the structure observed in the one-spot modeling is recovered by the two-spot approach, even though the spots on the primary
and secondary component occasionally exchange roles in the model.
Before epoch~0, the linear longitudinal relation for the secondary spot is not recovered, which we attribute to cross-talk
with the primary spot component and the comparably weak temperature contrast of the secondary spot there.
During epochs~0 to 20, the primary spot remains at a longitude of
about $30^{\circ}$, after which the model favors a spot at a longitude of about $200^{\circ}$. Here, however, the improvement in
fit quality is much more moderate, meaning that introducing this component may not actually be justified.

Both our one- and two-spot models result in solutions showing a structure with a rather well-defined linear time evolution of
longitude. We
therefore set up a two-spot model in which we fix the longitude of the secondary spot according to Eq.~\ref{eq:lon} and
vary only the longitude of its primary counterpart as well as both dimming factors and the normalization constant. The result
is displayed in the bottom panel of Fig.~\ref{fig:twospotevo}. The resulting pattern is similar to that obtained without the longitude
constraint and the fit quality is not much worse (Fig.~\ref{fig:chievo}).  

\begin{figure}
    \includegraphics[width=0.49\textwidth]{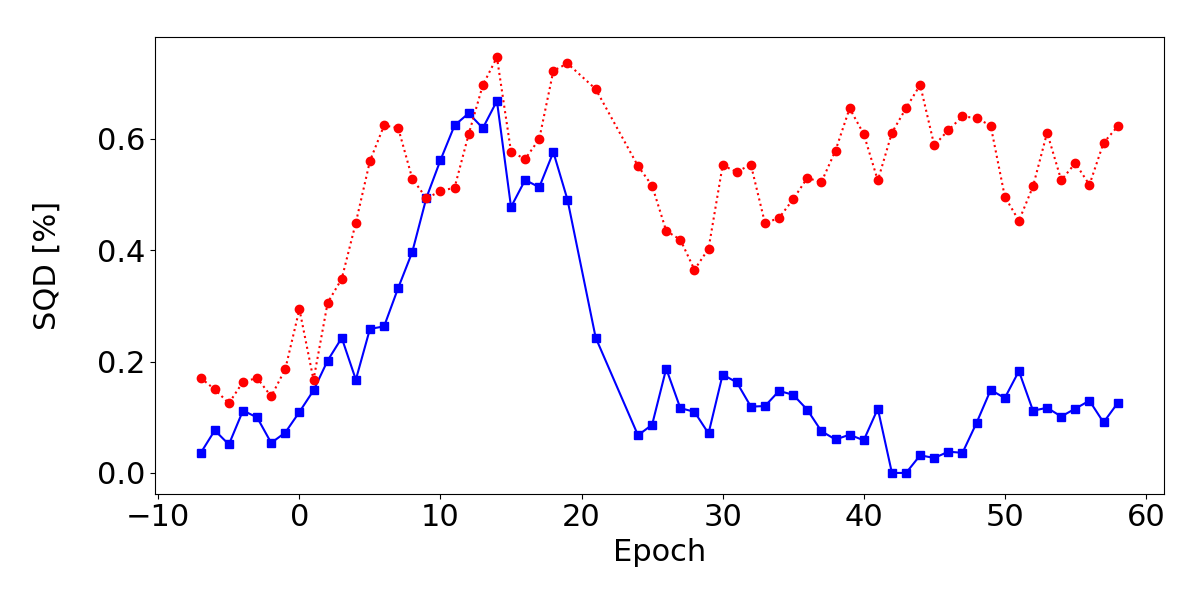}\\
    \caption{Spot quadrature depression (SQD) for the two-spot model with fixed
    longitudinal time evolution for spot on secondary. Red points correspond to spot on secondary
    and blue squares to spot on primary.  
    \label{fig:sqd2}}
\end{figure}

Based on the latter model, we calculate the SQD for both spots. The result is shown in Fig.~\ref{fig:sqd2}. The maximum SQD reached in
our modeling is about $0.6$\,\%, that is, the spot properties are such that the quadrature flux of the system is diminished by $0.6$\,\%.
While the secondary spot grows in influence before about epoch~10, after which the level is maintained, the primary spot shows
a shorter lifetime. According to our modeling, it grows and decays between epochs~0 and 20, corresponding to a lifetime of about $40$~d.
The actual structure observed here may be a comparably short-lived active region.
While the rate of growth appears to be similar for the primary and secondary spot, no decay is observed for the secondary spot. Again, the
interpretation of a primary spot at an SQD of approximately $0.1$\,\% is not entirely clear. This component could represent a real spot
or compensate inaccuracies in the modeling of the unspotted binary or starspot components.

\section{Discussion}

We modeled the light curve of the eclipsing binary \cone\ using the Nightfall code. First, we estimate the binary parameters
by a fit to the light curve of a single epoch, which we identified as particularly well suited for that purpose. Secondly, we study the starspot
evolution in the binary by modeling the \co\ light curves of all available epochs. In this process, we fix the binary parameters and only
adapt the starspot properties. In particular, we consider large spots with a fixed radius of $20^{\circ}$. While we adapt the longitude
and temperature contrast of the spots via their dimming factor, the spots remain fixed at the stellar equator. In our modeling,
we focus on models with one and two monolithic spots, which
capture the effect of longitudinally asymmetric starspot surface concentrations such as active regions or longitudes on the light curve. 

Both one- and two-spot models show the presence of at least
one spot moving in prograde direction at a rate of about $2.3^{\circ}$ per day.
At the beginning of the observed light curve, our two-spot model shows the emergence of a
new spot at a longitude of about $30^{\circ}$. This spot does
not strongly move in longitude. Its effect is significant for at least $20$~epochs or $40$~days. In our modeling, its contribution fades at
about epoch~25, after which the model favors a contribution from a spot at a higher longitude of about $200^{\circ}$. We caution here that some
impact of this spot might be absorbed in the model of the spot moving in prograde direction, which approaches in longitude. Also, the improvement in
fit quality obtained by introducing a second spot is rather moderate after epoch~20.

It appears that the lifetime of the spot
associated with the prograde motion is longer than the observed span of $145$~d.
Prograde spot motion has been observed by \citet{Heckert1995}, for example, in SS~Boo, however on a longer timescale.
In a study of light curves of short-term binaries observed
with Kepler, \citet{Balaji2015} found spots moving in prograde direction in $13$\,\% of the studied systems.
However, the major fraction of their binary sample show periods shorter than $2$~days.
In the short-period ($\approx 0.5$~d) eclipsing binary Kepler~11560447, \citet{Ozavci2018} find prograde shifts of active regions
in the K1IV primary component of that system. The authors derive a rate of shift of $2.4^{\circ}$~per day, which is quite compatible with
our result.

Different hypotheses can be invoked to explain the presence of two spots,
one moving in prograde direction and one with a rather constant
longitude, between epochs~0 and 20 \citep[see, e.g.,][]{Balaji2015}.
First, the spot moving in prograde direction may be located on a binary component, whose rotation is not synchronized with the binary orbit.
In this scenario, the other spot would have to be located on the other binary component, whose rotation would have to be synchronized.
The stellar rotation period of the component harboring the spot moving in prograde direction would then have to be shorter than the orbital
binary period by about $1.4$\,\% or $0.03$~d.
Second, the relative spot motion may be a result of differential rotation.
Assuming that both spots are really located on the same star, we can obtain an estimate of the strength of differential rotation on that star.
By making the extreme assumption that one spot is located at the equator and one at the pole, 
we can derive a lower limit for the absolute value of the ``relative horizontal shear'',
\begin{equation}
    \alpha \ge \left(\frac{P_2 - P_1}{P_1} \right) \; ,
\end{equation}
on that star \citep[e.g.,][]{Nagel2016}. This yields a lower limit of $0.014$ for $\alpha$ or $0.04$~rad per day for the absolute horizontal shear.
Clearly, the required latitudinal shear in rotation velocity must be larger to produce the same
effect if the spots are closer in latitude.  Also the sign of the shear remains unknown because we are ignorant of
the order of spot latitudes. The resulting shear is
consistent with the distribution of relative horizontal shear parameters derived by \citet{Reinhold2013} based on period analyses
of a large sample of Kepler light curves. In this scenario, the rotation of at least some latitude of the stellar surface has to
be synchronized with the binary orbit.
In a third, alternative scenario,
the observed behavior is caused by spot evolution, for example as a result of some sort of systematic evolution in the stellar
magnetic field,  resulting from an azimuthal dynamo wave for instance; this explanation was considered by \citet{Ozavci2018} for the case of Kepler~11560447.
Whether the small eccentricity in the system plays a role in any of these scenarios can only be studied in a larger sample of systems. 
While we cannot distinguish between the scenarios based on the current analysis, it shows the potential of the ever-growing number of
high-quality light curves of eclipsing binary stars for research on cool stars and stellar activity.

\section{Acknowledgements}

This work is based on data from the COROT Archive at CAB.
This research has made use of the ExoDat Database, operated at LAM-OAMP, Marseille, France, on behalf of the CoRoT/Exoplanet program. SC acknowledges support through DFG projects SCH 1382/2-1 and SCHM 1032/66-1.

\bibliographystyle{aa}
\bibliography{literature}

\end{document}